\documentclass{JAC2000}

\usepackage{graphicx}

\begin{document}
\title{HIGH CURRENT BEAM DYNAMICS IN AN ESS SC LINAC}

\author{M. Pabst, K. Bongardt, Forschungszentrum J\"ulich GmbH, Germany\\
A. Letchford, RAL, Didcot, U.K.}

\maketitle
\vspace{0.5cm}

\begin{abstract} 
Three alternative designs of the European Spallation Source (ESS) high energy linac are
described. The
most promising ones are either a normalconducting (nc) coupled cavity linac (CCL) up to
final energy or a change at 407 MeV  to only one group of 6 cell
superconducting (sc)  elliptical cavities.
\par
   Fully 3d Monte Carlo simulations are presented for both options,
optimized for reduced halo formation at the linac end. For the error free
matched case, especially  the halo formation in the longitudinal plane is
more pronounced for the hybrid solution with its superconducting
cavities, caused by the unavoidable phase slippage, but still quite well
acceptable for loss free ring injection. Simulations however  for a 30\%
mismatched dense core, surrounded in addition by 1.5\% halo particles are
showing few particles with very large amplitudes even in real space. This
case represents halo formation in front to end simulations, caused by
current fluctuations, filamented RFQ output distribution and enhanced by
accumulated field errors.

\end{abstract}

\section{Options for the High Energy Part of the 6\% d.c. ESS Linac}
The current reference design of the ESS contains a 1.334 GeV $H^{-}$ linac with a
5\% duty cycle, a 50 Hz repetition rate and a peak current of 114 mA. The
beam current is chopped with a 70\% duty cycle \cite{ref3}. The radio frequency is
280MHz for the two front end RFQs and DTLs. After funneling at 20 MeV final
acceleration to 1.334 GeV is accomplished in a nc CCDTL and
CCL operating at 560 MHz. A sc version of the high energy linac
is also being studied. The 1 msec long linac pulse is injected into 2
compressor rings, to produce a final beam pulse length of 1 $\mu$sec.

Any design of the ESS high energy linac must ensure loss free ring
injection. This demands an unfilamented 6d phase space distribution for the
linac beam.

Table 1, lists 3 high energy, 6\% duty cycle, linac design options for a 107
mA, 60\% chopped beam using 700 MHz structures. These were the linac
parameters from the ESS study \cite{ref1}.  The 700 MHz linac frequency is
also the same as considered for the CONCERT \cite{ref2} multi-user facility. The
following conclusions are valid for both 560 and 700 MHz frequencies, but
they are limited to linacs with about 6\% duty cycle.

\par

\begin{table}[htb]
\begin{center}
\caption{Options for the high energy part of the ESS 6\% d.c., 64 mA pulse current,
         700 MHz linac}
\begin{tabular}{|p{1.5cm}|p{1.4cm}|p{1.6cm}|p{2.0cm}|}
\hline
             & Normal \newline conducting \newline (nc) linac  &  Super-
\newline conducting \newline (sc) linac &
Hybrid solution  \\ \hline
Energy range & 105 -1334 MeV: CCL & 120-1334 MeV: \newline 
 $\beta=$0.52, \newline 0.65, 0.8 & 105-407 MeV: CCL \newline {$>$407 MeV:} 
$\beta=0.8$, sc \\ \hline
Total length & 631 m & 493 m & 148 m (nc) \newline + 267 m (sc) = 415 m  \\ \hline 
\# of cavities & 232 & 212 & 62 (nc) \newline + 116 (sc) = 178 \\ \hline
\# of klystrons & 116 & 212 & 31 (nc) \newline + 116 (sc) = 147 \\ \hline
Peak RF power per klystron &  2 MW & 0.4 MW, \newline 0.75 MW & 2 MW (nc), \newline 0.75 MW (sc) \\ \hline
Total peak RF Power& 232 MW & 101 MW & 137 MW \\ \hline
\# of circulators  & NONE & 212 & 116 \\ \hline
Cryogenic power & NONE &  4 MW & 3 MW \\ \hline
\end{tabular}
\end{center}
\end{table}

      For all three options, a doublet focusing system  with warm quadrupoles is
      assumed  either after 2   nc cavities
      \cite{ref1} or after  ( 2, 3, 4 ) 6-cell
      elliptical  sc cavities. The accelerating gradient
      is  kept  constant at $E_{o}T$ = 2.8 MV/m  for the nc cavities resp.
      $E_{o}$ = ( 5 MV/m, 8.50 MV/m, 13.7 MV/m ) for the 3 sc cavities.
      Two power coupler/ cavity are assumed for the $\beta=0.8$ sc cavities.
      The synchronous phase is kept constant at $-25^{o}$ in the nc  cells,
       whereas only the sc cavity midphase can be kept constant at $-25^{o}$ as a
       consequence of phase slippage. 
        All sc cavities   are assumed to  be  made out of 6 identical cells.
        The  relative $\beta$ dependence  of the transit time factor  is the same
        as for the SNS 805 MHz sc high  $\beta=0.76$ 6-cell cavities \cite{ref4},
        obtained from superfish calculations where end field  effects are
         included. The average transit time factor 
is smaller by at least 10\%  than the
      $\pi /4=0.79$ value  of   $\beta=1$ sc  elliptical cavites.
The average synchronous phase per cavity is smaller than $-37^{o}$ 
at beginning resp. end of each sc section.
      \par

       It is obvious from table 1, that a pure nc ESS linac version from
      105 MeV on is the cheapest in capital cost, but not in operating cost.
      A sc ESS linac from 120 MeV on requires much less peak RF power, but it
     is  in capital cost  quite expensive  and substantial R \& D  is necessary
      for the 50 Hz  pulsed mode behaviour  of  $\beta=0.52$ elliptical sc
cavities \cite{ref5, ref6}, including the ESS  2.7 msec long pulse option \cite{ref1}.
The hybrid solution with its two couplers/cavity is the shortest and the cheapest  one 
  for
capital plus 20 years operating time cost. By having only one
coupler/cavity  the ESS hybrid linac will be longer than the corresponding
nc one. Detailed pulsed power tests with 2 couplers/cavity are foreseen for
the 500 MHz, $\beta=0.75$ sc cavity teststand at FZ J\"ulich \cite{ref7}. The open
questions  are halo formation  at the end of
the ESS hybrid linac, resulting from the phase slippage and
enhanced by mismatch.

\section{Multiparticle Results for the Error Free Matched ESS Linac}

  In Fig. 1, 2 results  from  Monte Carlo simulations are shown for the
ESS nc linac at 105 MeV injection and at the 1334 MeV final energy. All
simulations are done with 10000 fully in 3d interacting particles. The 700
MHz bunch current is 107 mA, the normalized rms emittances are $0.3 \pi$ mm
mrad resp. $0.4 \pi^{o} MeV$.  The ratio between the full and zero current
tune is greater than ( 0.6, 0.5 ) transversely resp. longitudinally.  The
ratio between the transverse and longitudinal temperature in the rest
system is 0.66 at injection and about 1.3 at the linac end. All zero
current tunes are below $90^{o}$.  The rms radii are about 3 mm at injection
resp. 2 mm at  the final energy.
 \par
 The results in Fig. 1, 2 are obtained for the error free matched case. The
upper row corresponds to an  6d waterbag input distribution
limiting each
particle coordinate to $\sqrt 8$ of its rms value. As the ESS high
intensity nc linac is designed to avoid all kind of instabilities, driven
either by high space charge, temperature anisoptropy or resonance crossing
\cite{ref8}, almost no halo formation is visible at the linac end: there are no
particles outside $20 \epsilon_{rms}$ at the linac end. The  rms emittances
are changing by less than 10\%.
 \par

\begin{figure}[htb]
\centering
\includegraphics*[width=80mm]{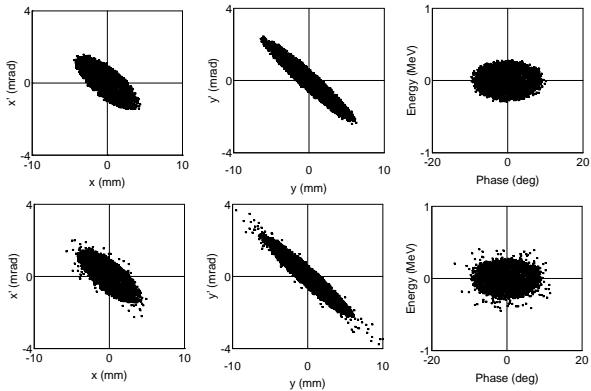}
\caption{Input distribution for the ESS linac with matched input. Upper row without,  
 lower row with 1.5\% initial halo}
\label{l2ea4-f1}
\end{figure}

\begin{figure}[htb]
\centering
\includegraphics*[width=80mm]{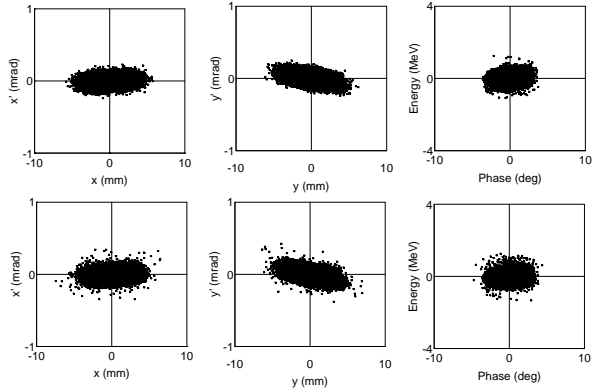}
\caption{Output distribution for the nc linac for matched input. Upper row without,
 lower row with 1.5\% initial halo}
\label{l2ea4-f1}
\end{figure}

   In the lower row 1.5\% halo particles are placed initially outside
the dense core  at  the surface of a  6d phase space boundary with $16 \epsilon_{rms}$. 
Each particle coordinate is now  limited to 4 times its rms value and there are
in each 2d phase space projection  less than $10^{-2}$ particles outside 
$10 \epsilon_{rms}$. About 1\%  halo particles above $10 \epsilon_{rms}$ are found in
simulations of  the 2.5 MeV  ESS RFQ  \cite{ref9}. Phase space correlations
between halo particles of a bunched beam in a periodic focusing channel are
reported  before \cite{ref8,ref10}.  At the ESS nc  linac end  there are now about
$10 ^{-3}$ particles outside $20 \epsilon_{rms}$, going up in phase space to about
$40 \epsilon_{rms}$.  But still all particles are limited to $\pm 10$ mm at the linac end
which is less than  half of the assumed 22 mm aperture radius. There are
less than $10^{-3}$ particles outside $\pm 1$ MeV.
  \par

   In Fig. 3 the output distributions are shown for the error free matched
ESS hybrid linac. Again the upper row is assuming an initial  6d waterbag
distribution without halo, whereas  the lower row  assumes a  distribution with
1.5\% halo.
  For a constant transverse full current tune of $45^{o}$ in the sc cavity section
the ratio between
the full and zero current tune is greater than ( 0.6 , 0.67 ) transversely
resp. longitudinally.  The ratio between the transverse and longitudinal
temperature in the rest system is 0.36 at 407 MeV and about 1.1
at the linac end. All zero current tunes are below $90^{o}$.  The rms radii are
about 2 mm at  407 MeV  resp. 1 mm at  the final energy.
  \par

\begin{figure}[htb]
\centering
\includegraphics*[width=80mm]{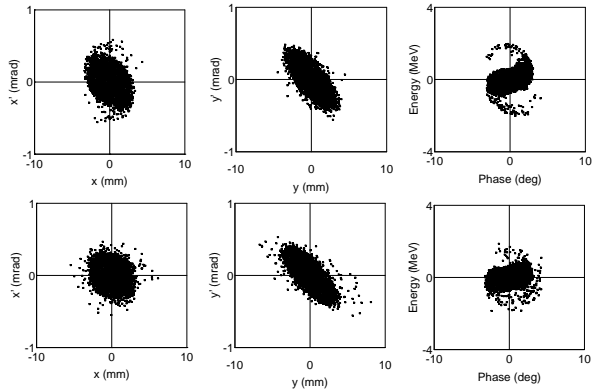}
\caption{Output distribution for the hybrid linac. Upper row without,
 lower row with 1.5\% initial halo}
\label{l2ea4-f1}
\end{figure}

 By comparing the nc output distribution in Fig. 2 with the hybrid one in
Fig. 3
 much more halo formation especially in the longitudinal plane  is visible
at the end of the ESS linac. Less  than $10^{-3}$ particles are
outside  $\pm 2$ MeV. The input distribution with initially
1.5\% halo particles  will lead to single particle amplitudes up to
9 mm at the ESS hybrid linac end, well outside the 6 mm  boundary
value of twice the  beam core size, predicted by particle-core models \cite{ref11}.
The reason is the phase slippage especially at beginning and end of the
sc section, where the beam velocity differs by $\pm 15\%$ from the cavity
design velocity. In the Monte Carlo simulations all particles are
experience the change of the synchronous phase from cell to cell in the
6-cell cavity. As a consequence, even the rms beam radii are oscillating
along the ESS hybrid linac, as the nc to sc 6d phase space matching is done
for the average cavity synchronous phase. 

\section{MONTE CARLO SIMULATIONS OF THE MISMATCHED ESS LINAC}
  In Fig. 4 phase space distributions are shown at the final 1334 MeV
   energy for the ESS nc  resp. hybrid linac, assuming  a mismatched input
distribution with 1.5\% initial halo particles. The upper row is showing the
nc and the lower row the  hybrid linac output  distributions.
  Excited is a pure high mode with 30
\% radial and 20\% axial mismatch of  the 3 bunch radii. The high
mode oscillation frequency of about $160^{o}$/period \cite{ref8} causes halo
formation in all 3 phase space planes as  single particle have  initial
frequencies of half the high mode oscillation frequency.

\begin{figure}[htb]
\centering
\includegraphics*[width=80mm]{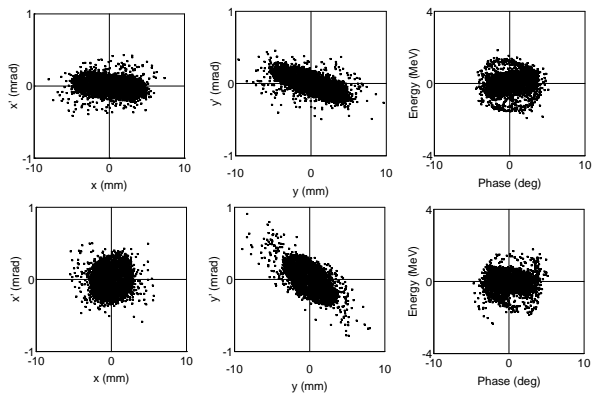}
\caption{Output distribution for the nc linac (upper row) and the hybrid linac (lower row).
The mismatched input distribution is surrounded by 
1.5\% initial halo.}
\label{l2ea4-f1}
\end{figure}

By assuming the same 30\% pure high mode excitation, but for an initial
distribution  without halo particles, the not shown  resulting phase space
distributions for both, the   nc  and the hybrid version of the ESS  high
intensity linac look quite similiar to the shown distribution in Fig. 4.
The only
difference are somewhat less particles nearby the bunch core.  But all
requirements  for loss free ring injection  are fullfilled: the final
normalized  transverse rms emittance is smaller than $0.4 \pi$ mm mrad  and
there are  about  $10^{-3}$ particles outside $20 \epsilon_{rms}$. Longitudinally there
are less than   $10^{-3}$ particles outside $\pm 2$ MeV , which is acceptable for
energy spread reduction by the bunch rotation system.
\par
Adding  initially 1.5\% halo particles to the 30\% mismatched dense core,
the linac output distributions  in  Fig. 4  show  a few particles withe
quite large amplitudes even in real space. Studies are going on for the
motion  of these particles in the linac to compressor ring transfer line ,
still effected by space charge forces \cite{ref1}.  Unconstrained hands on
maintenance requires less than 1 W/m uncollected deposited beam power at
linac end and at ring injection.This value corresponds to less than $10^{-7}/m$
uncollected particle loss for the ESS accelerator facility with its  5 MW
average beam power .
\par
 In a periodic focusing system  correlated field errors of $\pm 1\%$ even
for same limited periodes can cause noticable mismatch later on \cite{ref12}. In
front to end simulations of the complete ESS linear accelerator, including
the chopping and funnel lines, the halo formation at the linac end is
caused by current fluctuations, filamented RFQ output distribution and
enhanced by accumulated field errors. The so resulting halo formation  can
be estimated by assuming  30\% mismatch of an unfilamented beam, surrounded
by 1.5\% halo particles, at the entrance of the error free high energy
linac section.

\end{document}